\documentclass[%
reprint,
superscriptaddress,
amsmath,amssymb,
aps,
pra,
]{revtex4-2}

\usepackage{graphicx}
\usepackage{dcolumn}
\usepackage{bm}

\usepackage[english]{babel}

\usepackage{fancyhdr}

\begin{document}
\selectlanguage{english} 
\preprint{APS/123-QED}

\title{Sensitivity adjustable in-line high-temperature sensor based on metal microwire optical Fabry-Perot interferometer}

\author{Dewen Duan}
\email{ddw225@gmail.com}

\affiliation{School of Physics and Electronic Engineering, Sichuan University of Science and Engineering, No.1, Baita Road, Sanjiang New District, Yibin 644005, 
China}
\affiliation{School of Electronic and Information Engineering,Southwest University, No.2, Tiansheng Road, BeiBei District, Chongqing 400715, China}
\author{Zihao Zhao}
\affiliation{School of Electronic and Information Engineering,Southwest University, No.2, Tiansheng Road, BeiBei District, Chongqing 400715, China}
\author{Yi-Yuan Xie}
\email{yiyuanxie@swu.edu.cn}
\affiliation{School of Electronic and Information Engineering,Southwest University, No.2, Tiansheng Road, BeiBei District, Chongqing 400715, China}
%


\begin{abstract}
The optical fiber Fabry-Perot interferometer (FPI) has been widely investigated as a potential temperature sensor. To function as a temperature sensor, the cavity of the FPI is typically constructed from either silica fibers or polymers. The silica cavity FPIs can function at temperatures exceeding 1000\textdegree{}C. However, its temperature sensitivity is constrained by its relatively low thermal optical coefficient and thermal expansion of silica materials. Although the polymer cavity FPI exhibits a high temperature sensitivity, its cavity is susceptible to deterioration in high-temperature environments. Here, to overcome this challenge and achieve high-sensitivity temperature sensing in a high-temperature environment, we propose a new type of temperature FPI sensor by inserting and sealing a section of Cr20Ni80 metal microwire inside a section of silica hollow core fiber (HCF) spliced to standard single-mode fiber (SMF). The FPIs exhibit a high degree of temperature sensitivity due to the high thermal expansion of the Cr20Ni80 metal microwire. Since the Cr20Ni80 metal has a high melting temperature of 1400\textdegree{}C, such FPIs can function in high-temperature environments. Moreover, the temperature sensitivity of this FPI can be modified without affecting its reflection spectrum by changing the length of the metallic microwire situated within the hollow core fiber. The experimental results indicate that the proposed FPIs exhibit a temperature sensitivity of greater than -0.35nm/\textdegree{}C within the temperature range of 50\textdegree{}C to 440\textdegree{}C. Our proposed metal microwire-based FPIs are economical, robust, simple to fabricate, and capable of functioning in high-temperature environments, rendering them appealing options for practical applications. \\

\end{abstract}

\maketitle
\section{Introduction}
Temperature sensing, especially highly sensitive temperature sensing, is crucial in many application fields such as medical health, environmental monitoring, food production, biosensing, and aerospace. Currently, the most widely used sensors for monitoring temperature are electronic because they are massively scalable and easy to use. Unfortunately, these devices are difficult to multiplex, have relatively low resolution, and are not immune to electromagnetic interference. Optical fiber temperature sensors have been widely investigated as replacement temperature monitoring solutions due to their inherent advantages such as multiplexing capability, low mass, high resolution, biocompatibility, electromagnetic immunity, remote sensing capability, and absence of electrical signals at the measurement location\cite{Bao19, Polz16, Jia18, Gangwar23}. A considerable number of optical fiber temperature sensors have been proposed. These sensors may be loosely split into three types based on their diverse functioning principles: fluorescence-based optical fiber sensors, optical fiber grating-based sensors, and optical fiber interferometer-based sensors.  Fluorescence-based sensors utilize fluorescence materials attached to optical fibers\cite{Zeltner16, Bian21, Huang22}. Grating-based sensors include fiber Bragg gratings (FBGs) and long-period fiber gratings (LPFGs)\cite{Polz16, Xu24, Qiu24, Deng21}. Interferometer-based sensors include Michelson interferometers, Mach-Zehnder interferometers (MZIs), Fabry-Perot interferometers(FPIs), and Sagnac interferometers\cite{Han21, Deng17, Ding21, Jia18, Li19, Zhao20}. Other types of optical temperature sensors include the isopropanol-sealed optical microfiber couplers\cite{Zhao18}, surface plasmonics sensors\cite{Chang20}, Raman scattering in optical fibers\cite{Mizuno24}. Among the various temperature sensors, those based on fluorescence are not multiplexing-friendly. Grating-based sensors are typically characterized by high resolution and a wide dynamic measurement range, but their fabrication usually requires complex fabrication processes and expensive equipment. In contrast, interferometer-based sensors are often relatively simple in structure, straightforward to fabricate, and cost-effective.  

However, most optical fiber interferometer cavities are composed of silica glass, which has low thermal optical and thermal expansion coefficients. This results in a relatively poor temperature sensitivity, which restricts the application of interferometer-based sensors. Several techniques have been employed to enhance the temperature sensitivity of interferometer-based optical fiber sensors. One such technique is filling the cavity of the interferometer, which is formed by a hollow-core photonic crystal fiber, with a material of high thermal expansion and thermal optical coefficients, such as alcohol or polydimethylsiloxane (PDMS)\cite{Qian11, Gao18, Gao21}. Another technique involves the extension of the cavity of the interferometer to enhance the temperature effect accumulation length of the sensors, thereby increasing the temperature sensitivity of the sensor\cite{Li19}.
Nevertheless, despite the high thermal expansion and thermal optical coefficients of liquids and polymers, most of these materials cannot withstand temperatures above 100°C. This limitation restricts their application to below this threshold. Extending the cavity length of the interferometer will reduce the spatial resolution of the sensor. Here, we propose a high-sensitivity in-line optical fiber Fabry-Perot interferometer temperature sensor based on a metal microwire. The sensor is fabricated by inserting and fixing a section of Cr20Ni80 metal microwire in a silica hollow core fiber (HCF) spliced at the end of a single-mode fiber (SMF). Due to the high thermal expansion coefficient and the melting point of 1400\textdegree{}C in the Cr20Ni80 metal microwire, such a sensor has a high temperature sensitivity and can function in high-temperature environments. Furthermore, the metal microwire end can be polished to achieve a high light reflection coefficient, thereby enabling the formation of a fiber-optic Fizeau-type FPI sensor and a substantial increase in the cavity length. This allows such sensors to have a high multiplexing capability \cite{Rao05, Rao06}.    

\section{Sensor working principle}

\begin{figure}[ht!]
	\centering\includegraphics{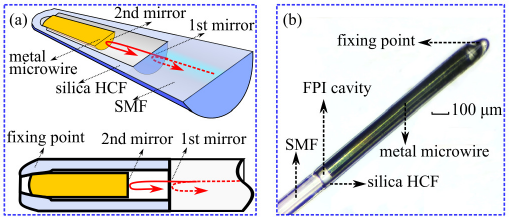}
	\caption{(a) Schematic of the metal microwire-based Fabry-Perot interferometer (FPI) fabricated by inserting and fixing a section of a metal microwire in a silica hollow core fiber (HCF) spliced at the end of a single-mode fiber (SMF), and (b) a microscopic photo of one fabricated metal microwire-based FPIs; the length of the metal microwire is approximately 1078 $\mu m$, and the length of the FPI cavity is approximately 88.5$\mu m $.}\label{fig:1}
\end{figure}

A principle schematic and microscopic photograph of a fabricated metal microwire-based FPI are presented in Fig.~\ref{fig:1}. The SMF end at its fusion junction with the silica HCF and the metal microwire end located near the end of the SMF formed the two mirrors of the FPI cavity. The section of the metal microwire located within the silica HCF, with one end situated at a considerable distance from the silica HCF end, was fixed by the collapse of the silica HCF, which was used as the temperature-sensitive component. The temperature change induced FPI cavity changes can be expressed as  
\begin{equation}
	\ \frac{\Delta l}{\Delta T} = \alpha_s l_s - \alpha_m l_m
	\label{eq:1}.
\end{equation}
where $l_m$ and $l_s$ are the lengths of the metal microwire and the silica HCF respectively. $\alpha_m$ and $\alpha_s$  are the linear expansion coefficients of the metal microwire and silica HCF, respectively. The original FPI cavity length (the gap between the SMF apex and the metal microwire end ) is $l_c$=$l_s$-$l_m$. Even though the length of the silica HCF $l_s$ is larger than the length of the metal microwire, $\Delta l$ is negative when the temperature change is positive since the linear expansion coefficient of the metal microwire $\alpha_m$ is much larger than that of the silica HCF $\alpha_s$. This means that the FPI cavity length decreases as the temperature increases. Eq.(\ref{eq:1}) can also be rewritten in the following form 

\begin{equation}
	\ \frac{\Delta l}{\Delta T} = \alpha_sl_c + (\alpha_s - \alpha_m)l_m 
	\label{eq:2}.
\end{equation}
Eq.(\ref{eq:2}) shows that when the cavity length of the FPI $l_c$ is certain, both choosing a metal microwire with a high linear expansion coefficient and increasing the metal microwire length $l_m$ will enhance the temperature sensitivity of the FPI. For example, since aluminum has a greater linear expansion coefficient than copper, the same length of aluminum microwire inserted in the silica HCF will result in a higher temperature sensitivity than that of the copper microwire. One advantage of our proposed metal microwire-based FPI is that the temperature sensitivity of the sensor can easily be enhanced without affecting the cavity length of the FPI and the reflection signal quality by increasing the length of the metal microwire $l_m$. The disadvantage of increasing the length of the metal microwire $l_m$ is that it decreases the spatial resolution of the sensor. 

\section{Experimental results and discussion} 

\begin{figure}[ht!]
	\centering\includegraphics{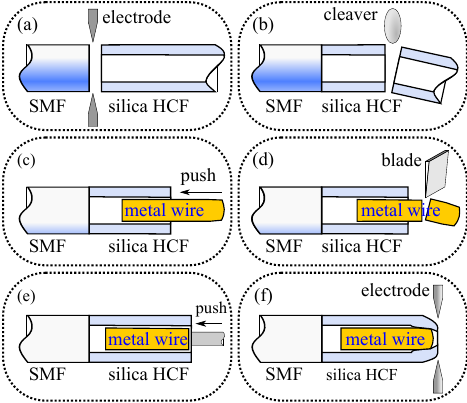}
	\caption{ Fabrication processing of the metal microwire-based FPIs.}\label{fig:2}
\end{figure}

The detailed fabrication procedure of the metal microwire-based FPI sensor is shown in Fig.~\ref{fig:2}. First, a section of silica HCF was fusion-spliced to the end of a section of SMF and cleaved with a designed length spliced to the SMF (Fig.~\ref{fig:2} (a) and (b)). In the splicing process, the fusion discharge current is only approximately half that of the program for splicing an SMF to another SMF, and the silica HCF is approximately 20 $\mu m $ away from the electrode pair center to prevent the silica HCF from collapsing and deforming under fusion discharge. Second, a section of the metal microwire was inserted into the silica HCF and trimmed to a length designed to remain within the silica HCF (Fig.~\ref{fig:2} (c) and (d)). The length of the retained section of the metal microwire should be less than that of the silica HCF. Third, the section of the metal microwire was further pushed to be completely inside the silica HCF with a designed distance (the FPI cavity length) from the SMF end by an optical fiber taper (Fig.~\ref{fig:2} (e)). This was followed by the metal microwire being fixed by collapsing and deforming the silica HCF end under fusion discharge (Fig.~\ref{fig:2} (f)). To ensure that the metal microwire was securely fixed within the silica HCF, the metal microwire was inserted into the silica HCF for a short distance from the intended point of collapse and deformation of the silica HCF. Fig.~\ref{fig:1} (b) shows a microscopic photograph of one fabricated metal microwire-based FPI.

\begin{figure}[ht!]
	\centering\includegraphics{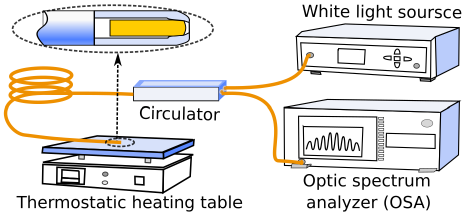}
	\caption{FPI sensor temperature performance investigation setup.}\label{fig:3}
\end{figure}

\begin{figure}[ht!]
	\centering\includegraphics{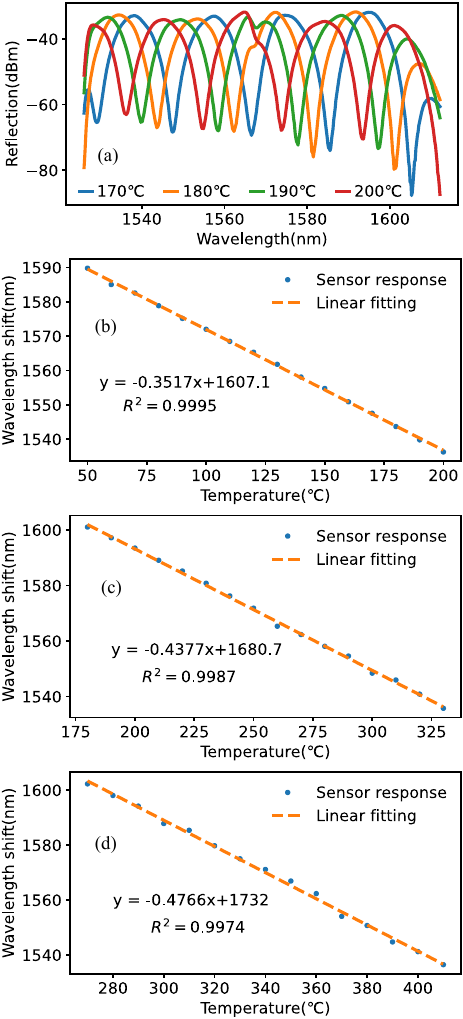}
	\caption{(a) The $\sim$88.5 $\mu m $ long metal microwire-based FPI reflection spectrum blue shifts with the temperature increase. (b), (c) and (d) Resonant dip wavelength shifts as a function of temperature from 50\textdegree{}C to 440\textdegree{}C; the microscopic image of the FPI is shown in Fig.~\ref{fig:1}(b). }\label{fig:4}
\end{figure}

\begin{figure}[ht!]
	\centering\includegraphics{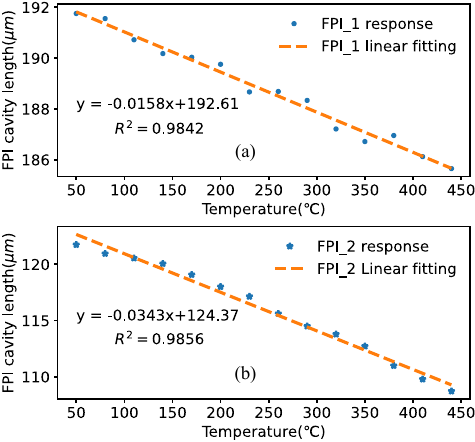}
	\caption{The cavity lengths of the two microwire-based FPIs decrease with increasing temperature. Their metal microwire lengths are:  (a) FPI\underline{ }1 is approximately 988 $\mu m $ and (b) FPI\underline{ }2 is approximately 2110 $\mu m $. }\label{fig:5}
\end{figure}

The metal microwire-based FPI sensor temperature performance investigation setup is depicted in Fig.~\ref{fig:3}. The broadband light from the white light source was launched into the metal microwire-based FPI sensor through the optical fiber connected to the optical circulator. The reflection spectrum of the FPI was recoupled to the optical circulator and recorded using an optical spectrum analyzer (OSA, Anritsu, MS9740A). The metal microwire-based FPI sensor was mounted on a thermostatic heating table and heated from 50\textdegree{}C to 440\textdegree{}C with an incremental step of 10\textdegree{}C. The temperature response of a microwire-based FPI with a length of approximately 88.5 $\mu m $ (Fig.~\ref{fig:1}(b)) is presented in Fig.~\ref{fig:4}. Fig.~\ref{fig:4}(a) illustrates that the reflection interference spectrum shifts to the short-wavelength side (blue shifting) as the temperature increases from 170\textdegree{}C to 200\textdegree{}C.  Due to the limited bandwidth of our white light source, we were only able to monitor the FPI sensor's temperature response within an approximate range of 150\textdegree{}C. Consequently, we identified three resonance dips of the FPIs as the temperature increased from 50\textdegree{}C to 410\textdegree{}C. The results of these observations are presented in Fig.~\ref{fig:4}(b), (c) and (d). The results indicated that this sensor's reflection spectrum resonant dips shift to shorter wavelengths with increasing temperature, and their sensitivity is greater than -0.35 nm/\textdegree{}C. . As illustrated in Fig.~\ref{fig:4}, 
the high temperature sensitivity of the metal microwire-based FPI results in a significant shift in the resonant wavelength when monitoring temperatures. Consequently, any fringe valley would shift outside the wavelength range of the spectrometer when attempting to cover a broad temperature range. To overcome this limitation, two techniques can be employed: the first is to stitch several resonant valleys together, as  demonstrated by Liu \textit{et al.}\cite{Liu:17}, and the second is to directly demodulate the temperature-induced FPI cavity length change, as illustrated in Fig.~\ref{fig:5}.

Additionally, two microwire-based FPIs were fabricated with various lengths of metal microwire and experimentally compared for their respective cavity length decrease as the temperature increases from 50\textdegree{}C to 440\textdegree{}C, with an incremental step of 30\textdegree{}C. The results of the experiment are presented in Fig.~\ref{fig:5}(a) and (b). Fig.~\ref{fig:5} (a) demonstrates that the approximately 192 $\mu m $  FPIs (with a metal microwire length of roughly 988 $\mu m $) exhibit a sensitivity of -0.0158 $\mu m$/\textdegree{}C (Fig.~\ref{fig:5} (a)). In contrast, the approximately 124 $\mu m $  FPI (with a metal microwire length of roughly 2110 $\mu m $) exhibited a sensitivity of -0.0343 $\mu m $/\textdegree{}C (Fig.~\ref{fig:5} (b)). The values obtained from the product data sheet for the linear thermal expansion coefficient of Ni80Cr20 microwire 18x 10$^{-6}$ were input into Eq. Eq.(\ref{eq:1}), resulting in the determination that the silica HCF we utilized has a linear thermal expansion coefficient of 1.68 x 10$^{-6}$ for the short metal microwire-based FPI and 1.65 x 10$^{-6}$ for the long metal microwire-based FPI, respectively. The discrepancy in the calculated values is primarily attributable to the lack of precision in measuring the length of the metal microwire. Compared with the values reported in the literature, the values are larger than those reported for fused silica but smaller than those reported for many optical glasses\cite{Baak69, Deng20}. Fig.~\ref{fig:5} also demonstrates that the cavity length decrease rate of the longer metal microwire-based FPIs is faster than that of the shorter metal microwire-based FPIs as the temperature increases. The rates of decrease in cavity lengths for the longer metal microwire-based FPI and the shorter metal microwire-based FPIs are 0.0343 $\mu m $/\textdegree{}C (FPI\underline{ }2 in Fig.~\ref{fig:5} (b)) and 0.0158 $\mu m $/\textdegree{}C (FPI\underline{ }1 in Fig. Fig.~\ref{fig:5} (a)), respectively. These values are in accordance with the predictions of Eq. 2. Moreover, the outcomes of this experiment indicate that an increase in the length of the metal microwire used in a metal microwire-based FPI will increase the temperature sensitivity of the sensor.

In this study, the Cr20Ni80 metal microwire was employed as the temperature-sensitive component of the FPI device. 
Indeed, a variety of alternative metal microwires may be selected for this purpose, including gold wire, platinum wire, copper wire, aluminum wire, and steel wire. The Cr20Ni80 metal microwire was ultimately selected due to its advantageous properties compared to other metal wires. The Cr20Ni80 wire displays a relatively high thermal expansion coefficient of 18×10$^{-6}$, although this value is lower than that of lead and aluminum wires. Its melting point is considerably higher than that of lead and aluminum wires. Moreover, the melting point of the metal microwire exceeds the softening point of silica glass fiber, which results in the FPI's damage temperature threshold extending to the softening point of the silica glass fiber. Furthermore, it is more cost-effective compare to gold wire and platinum wire.

Unlike the silica material, the metal wire may absorb specific directions and wavelengths of external electromagnetic fields, resulting in a thermal effect that affects the accuracy of temperature measurements. However, given the small and short dimensions of the metal wire, the heat generated by our typical exposure to external electromagnetic fields can be considered negligible. Nevertheless, the wire does exhibit a capacity to absorb shorter wavelength electromagnetic fields, including light, infrared, and far infrared radiation. The underlying mechanisms involved are complex and warrant further investigation. 


\section{Conclusion}

We conceptually proposed a temperature sensitivity adjustable metal microwire-based FPI. This FPI is fabricated by inserting and sealing a section of Cr20Ni80 metal microwire inside a silica hollow core fiber (HCF) spliced into standard single-mode fiber (SMF). Theoretically, the sensor can function in a high-temperature environment, with a temperature slightly below the softening point of silica optical fiber.  Moreover, the temperature sensitivity of the sensor can be enhanced by increasing the metal microwire situated and fixed within the silica HCF. A temperature investigation of an FPI with a cavity of $\sim $88.5 $\mu m $, fabricated by insertion and fixation of a $\sim $1078 $\mu m $ metal microwire inside a section of silica HCF, indicates that such a sensor exhibits a sensitivity of over -0.35 nm/\textdegree{}C, as evidenced by the tracing of the resonant dip of the reflection spectrum within the temperature range of 50C and 400C. Furthermore, a comparison was conducted between two FPI devices with disparate cavity lengths and metal microwire lengths. The findings demonstrated that an increase in the length of the metal microwire would result in an augmented temperature-induced FPI cavity length, thereby enhancing the device's temperature sensitivity. The proposed metal microwire-based FPI is an economically viable, robust, straightforwardly manufactured, and high-temperature tolerant device that can function in challenging environments, making it an appealing option for practical applications.

\vspace{0.5cm}

\textbf{Acknowledgements}  We wish to thank Prof. Yinhong Liao for his assistance in the sensor's temperature performance investigation experiments. 

\textbf{Founding} This work is supported by the Chongqing Ph.D. “Through Train” Scientific Research Project (Grant Number: sl202100000172). 

\textbf{Conflict of interest} The authors have no competing interests to declare that are relevant to the content of this article.

\textbf{Data Availability Statement} Data underlying the results presented in this paper are not publicly available at this time but may be obtained from the authors upon reasonable request.

\textbf{Author's Contributions} D.Duan designed the study, performed the experiments, analyzed the data, and drafted the manuscript. Z.Zhao participated in the experiments and data analysis. Y.Xie participated in the study design and coordination and helped draft the manuscript. All authors read and approved the final version of the manuscript.

\bibliography{reference}

\end{document}